\documentclass[]{aastex631}

\usepackage{gensymb}

\graphicspath{{./}{figures/}}

\begin{document}

\title{Quick-Look Pipeline Lightcurves for 9.1 Million Stars Observed Over the First Year of the TESS Extended Mission}

\correspondingauthor{Michelle Kunimoto}
\email{mkuni@mit.edu}

\author[0000-0001-9269-8060]{Michelle Kunimoto}
\affiliation{Kavli Institute for Astrophysics and Space Research, Massachusetts Institute of Technology, Cambridge, MA 02139}

\author[0000-0003-0918-7484]{Chelsea Huang}
\affiliation{Kavli Institute for Astrophysics and Space Research, Massachusetts Institute of Technology, Cambridge, MA 02139}

\author[0000-0002-5308-8603]{Evan Tey}
\affiliation{Kavli Institute for Astrophysics and Space Research, Massachusetts Institute of Technology, Cambridge, MA 02139}

\author[0000-0003-0241-2757]{Willie Fong}
\affiliation{Kavli Institute for Astrophysics and Space Research, Massachusetts Institute of Technology, Cambridge, MA 02139}

\author[0000-0002-2135-9018]{Katharine Hesse}
\affiliation{Kavli Institute for Astrophysics and Space Research, Massachusetts Institute of Technology, Cambridge, MA 02139}

\author[0000-0002-1836-3120]{Avi Shporer}
\affiliation{Kavli Institute for Astrophysics and Space Research, Massachusetts Institute of Technology, Cambridge, MA 02139}

\author[0000-0002-5169-9427]{Natalia Guerrero}
\affiliation{Department of Astronomy, University of Florida, Gainesville, FL 32611}

\author[0000-0002-9113-7162]{Michael Fausnaugh}
\affiliation{Kavli Institute for Astrophysics and Space Research, Massachusetts Institute of Technology, Cambridge, MA 02139}

\author[0000-0001-6763-6562]{Roland Vanderspek}
\affiliation{Kavli Institute for Astrophysics and Space Research, Massachusetts Institute of Technology, Cambridge, MA 02139}

\author[0000-0003-2058-6662]{George Ricker}
\affiliation{Kavli Institute for Astrophysics and Space Research, Massachusetts Institute of Technology, Cambridge, MA 02139}

\begin{abstract}
We present a magnitude-limited set of lightcurves for stars observed over the TESS Extended Mission, as extracted from full-frame images (FFIs) by MIT’s Quick-Look Pipeline (QLP). QLP uses multi-aperture photometry to produce lightcurves for $\sim$1 million stars each 27.4-day sector, which are then searched for exoplanet transits. The per-sector lightcurves for 9.1 million unique targets observed over the first year of the Extended Mission (Sectors 27 -- 39) are available as High-Level Science Products (HLSP) on the Mikulski Archive for Space Telescopes (MAST). As in our TESS Primary Mission QLP HLSP delivery \citep{Huang2020}, our available data products include both raw and detrended flux time series for all observed stars brighter than TESS magnitude $T = 13.5$, providing the community with one of the largest sources of FFI-extracted lightcurves to date.

\end{abstract}


\section{Introduction}

In July 2020, NASA's Transiting Exoplanet Survey Satellite \citep[TESS;][]{Ricker2015} completed its two-year Primary Mission to search for transiting exoplanets around nearby bright stars. TESS' four cameras observed $96\degree\times24\degree$ sectors of the sky for 27.4 days each, totalling 26 sectors over the Primary Mission. Overall, $\sim$200,000 pre-selected stars received 2-minute cadence observations, which were processed by the TESS Science Processing Operations Center \citep[SPOC;][]{Jenkins2016} pipeline. However, TESS also recorded measurements of its entire field of view in 30-minute sampled full-frame images (FFIs), enabling the flux measurements of tens of millions of stars. Between 2- and 30-minute observations, the TESS Primary Mission resulted in the identification of 2,241 planet candidates \citep{Guerrero2021}.

TESS' current Extended Mission is capturing millions more stars in FFIs, both increasing the observation baseline for stars re-observed from the Primary Mission and increasing the the satellite's overall coverage of the sky. FFI cadence was also reduced from 30 to 10 minutes. Relevant for exoplanet searches, the faster cadence better resolves transit shapes and improves the detectability of short-duration signals. Here, we present a magnitude-limited set of lightcurves extracted from Extended Mission FFIs each TESS sector by MIT's Quick-Look Pipeline \citep[QLP;][]{Huang2020}. We report on the delivery to the Mikulski Archive for Space Telescopes (MAST) of these High-Level Science Products (HLSPs) for 9,115,386 targets observed in the southern ecliptic hemisphere (Sectors 27 -- 39). HLSPs for the northern hemisphere and ecliptic (Sectors 40 -- 55) are upcoming.

\section{Lightcurve Production}

The full QLP lightcurve extraction and post-processing procedure is described in \citet{Huang2020}. In summary, QLP first calibrates FFIs using the TICA software \citep{Fausnaugh2020}, updated to be compatible with 10-minute cadence frames, to correct for instrumental effects. Significant levels of scattered light are then removed by subtracting a global background with nebuliser \citep{Irwin1985}, and images are median-subtracted using a median combination of 40 good quality images identified within the TESS orbit. An astrometric solution is calibrated with bright, un-saturated stars (TESS magnitude between $8 < T < 10$) to identify the locations of each star brighter than $T = 13.5$ mag in the TESS Input Catalog \citep[TIC;][]{Stassun2019}. QLP then performs multi-aperture photometry by summing the flux contained in five circular apertures centred on these positions, ranging from 1.75 to 8 pixels in radius. A local background is subtracted to remove small-scale features, and this difference flux is converted into an absolute flux based on what is expected given a target's TESS magnitude. Finally, the resulting lightcurves are detrended using basis spline fits, and poor quality datapoints are flagged.

\begin{figure}
    \centering
    \includegraphics[width=0.43\textwidth]{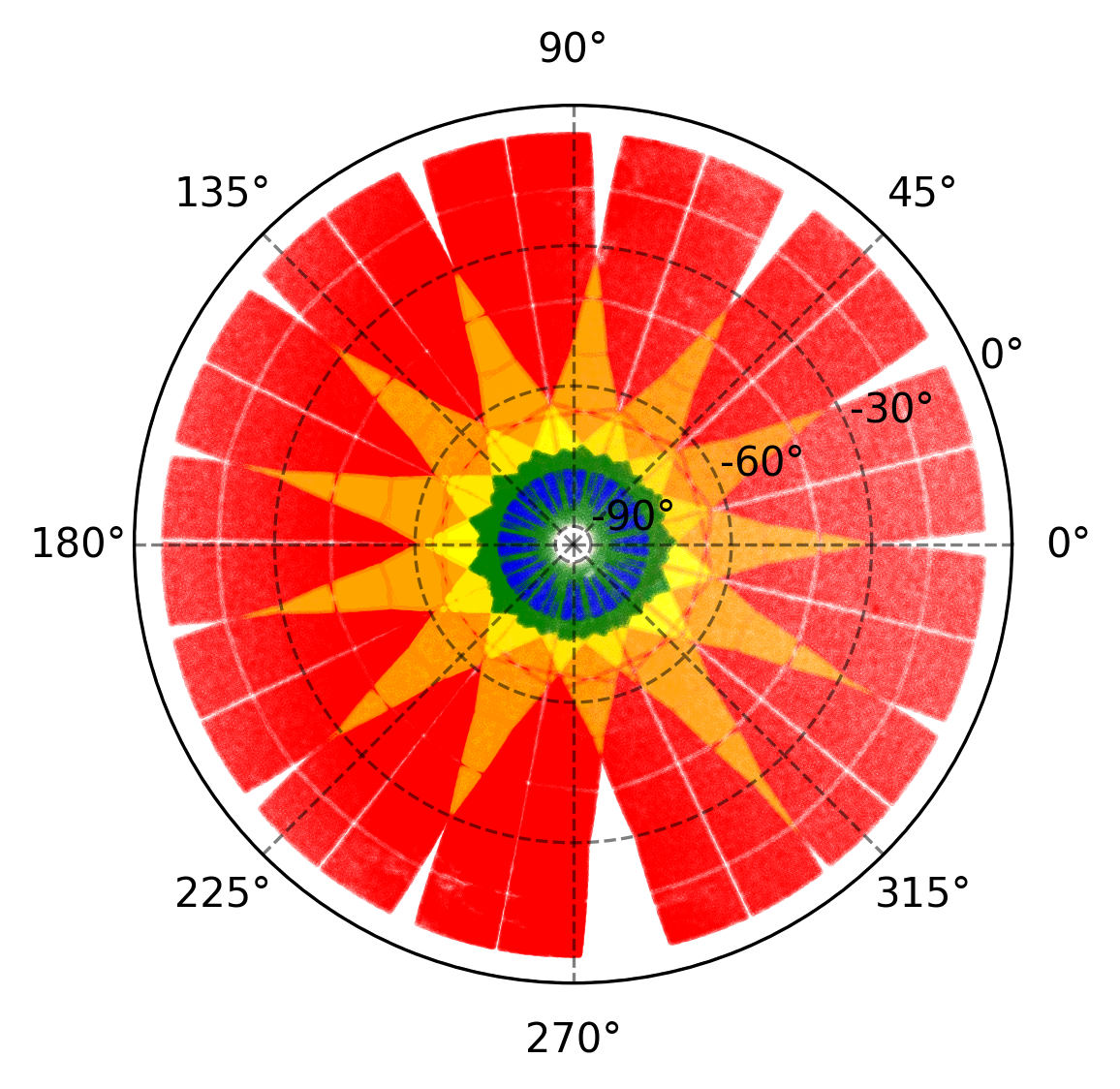}
    \includegraphics[width=0.55\textwidth]{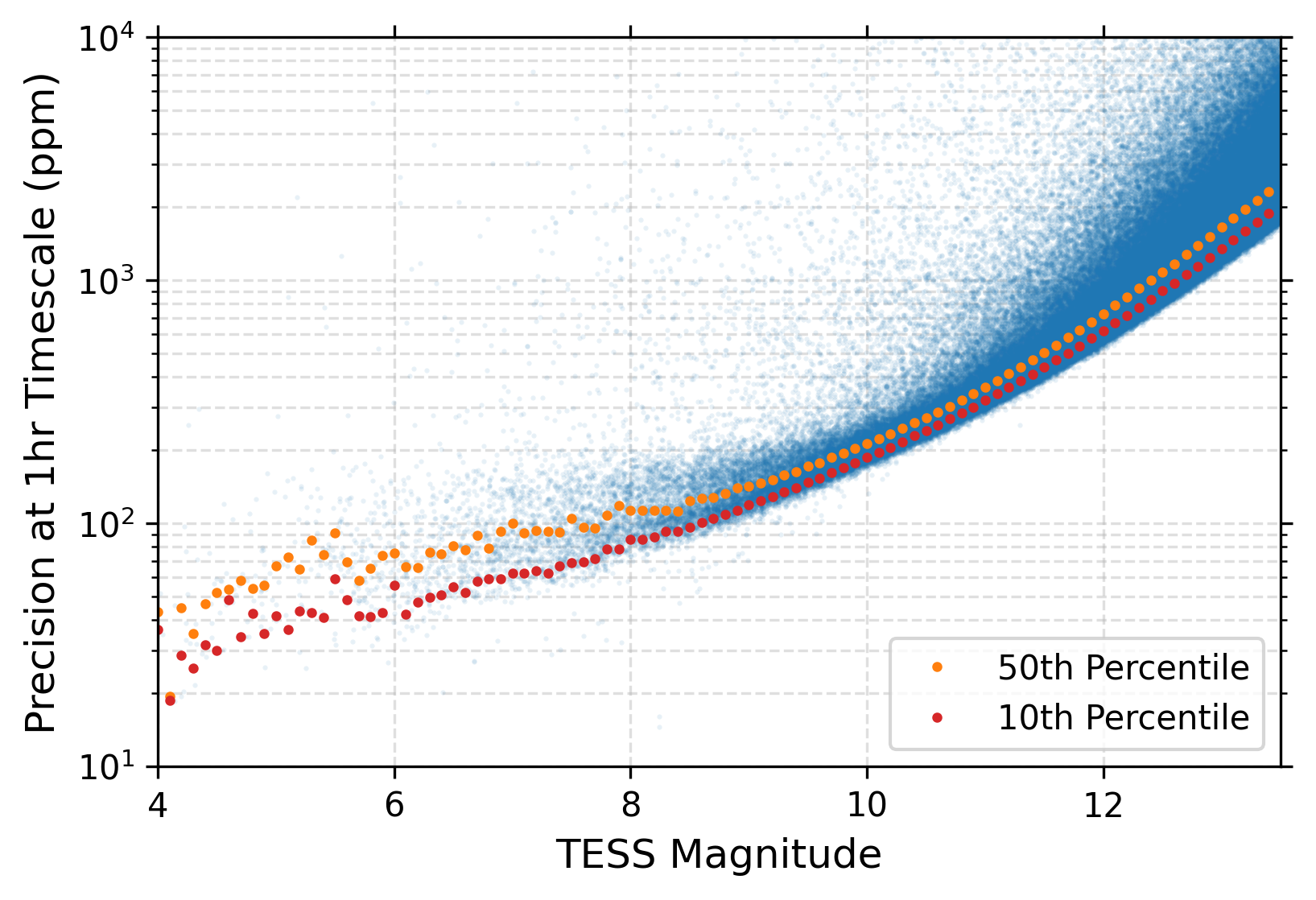}
    \caption{\textbf{Left:} A polar projection of the ecliptic coordinates of targets with QLP lightcurves from the southern ecliptic hemisphere (Sectors 27 -- 39). Targets are colour-coded by the number of observed sectors, either 1 (red), 2 (orange), 3 (yellow), 4 -- 12 (green), or 13 (blue). \textbf{Right:} 1hr precision of 16,000 randomly selected lightcurves from each orbit across Sectors 27 -- 39. The 50th and 10th percentiles of the precision binned every $\Delta T = 0.1$ mag are shown in orange and red, respectively.}
    \label{fig}
\end{figure}

QLP produces lightcurves for $\sim$1 million stars each sector, with the total Sector 27 -- 39 coverage shown in the left panel of Figure \ref{fig}. The right panel of Figure \ref{fig} shows the precision of 16,000 randomly selected detrended lightcurves from each orbit as a function of TESS magnitude, based on lightcurves made with a 2.5-pixel circular aperture. The precision is estimated as 1.48 times the median absolute standard deviation of the lightcurve, scaled to a 1hr timescale. The 50th and 10th percentiles of the precision binned every $\Delta T = 0.1$ mag are shown in orange and red, respectively. After the completion of a sector, each target's lightcurves are merged with all previous data available to that target. These multi-sector lightcurves are the primary data products used in the remainder of QLP processing, from the transit search to planet candidate report generation.

We used TICA v0.2.1, TIC v8.1 for Sectors 27 -- 38, and TIC v8.2 for Sector 39.

\section{High-Level Science Products}


The full description of QLP data products provided as HLSPs to MAST is given in the supplemental material of \citet{Huang2020}. As in the Primary Mission, we combine data from both orbits of each TESS sector for all observed targets, and deliver this data in the form of both FITS files and comma-separated ASCII txt files. Among other information, we provide the un-detrended flux from the optimal aperture in the FITS files under the keyword \textbf{SAP\textunderscore FLUX}, and detrended flux under \textbf{KSPSAP\textunderscore FLUX}. We determine the optimal aperture for each target based on its TESS magnitude. The flux measurements from relatively bigger/smaller apertures are under the keywords \textbf{KSPSAP\textunderscore FLUX\textunderscore LAG} and \textbf{KSPSAP\textunderscore FLUX\textunderscore SML}, respectively. The observation times, provided under the \textbf{TIME} keyword, are barycentric-corrected using J2000 coordinates. We also provide quality flags for each cadence under the \textbf{QUALITY} keyword. Cadences flagged as poor quality by QLP are assigned the bit value 4096, while other non-zero flags are adopted from SPOC FFI headers \citep{Jenkins2016} and combined in a bitwise operation. In the more light-weight ASCII files, we provide the time, un-detrended flux, and local background flux estimate, with all flagged cadences rejected.

We made the following updates for the Extended Mission HLSP FITS files:

\begin{itemize}
    \item Added \textbf{RA\textunderscore ORIG} and \textbf{DEC\textunderscore ORIG} headers, containing the J2015.5 right ascension and declination for each target. If not available, the values are set to -1.0.
    \item Provided the specific version of the TIC used for the production of the lightcurve and querying of the stellar parameters in the \textbf{TICVER} header.
    \item Changed the value of the \textbf{CALIB} header from ``TICA'' to ``TICA v0.2.1'' to reflect the specific version of TICA used.
\end{itemize}

\section{Acknowledgements}
We thank the entire TESS Mission team for years of effort to make this work possible. This paper includes data collected by the TESS mission, which are publicly available from the Mikulski Archive for Space Telescopes (MAST). The TESS mission is funded by NASA’s Science Mission Directorate.

\bibliography{refs}

\end{document}